\documentclass[a4paper,aps,11pt,preprint,showpacs]{revtex4}
\usepackage{graphicx,amsmath,revsymb,bm,amssymb,amsfonts,amsthm}
\begin{document}

\title{Single photoeffect on helium-like ions in the non-relativistic region}
\author{A.I.~Mikhailov$^{\,\mathrm{a,b}}$, A.V.~Nefiodov$^{\,\mathrm{a,c}}$,
G.~Plunien$^{\,\mathrm{c}}$}
\affiliation{$^{\mathrm{a}}$Petersburg Nuclear Physics Institute,
188300 Gatchina, St.~Petersburg, Russia \\
$^{\mathrm{b}}$Max-Planck-Institut f\"ur Physik komplexer Systeme,
N\"othnitzer Stra{\ss}e 38, D-01187 Dresden, Germany \\
$^{\mathrm{c}}$Institut f\"ur Theoretische Physik, Technische
Universit\"at Dresden, Mommsenstra{\ss}e 13, D-01062  Dresden,
Germany }

\date{Received \today}
\widetext
\begin{abstract}
We present a generalization of the pioneering results obtained for
single K-shell photoionization of H-like ions by M. Stobbe [Ann.
Phys. 7 (1930) 661] to the case of the helium isoelectronic
sequence. The total cross section of the process is calculated,
taking into account the correlation corrections to first order of
the perturbation theory with respect to the electron-electron
interaction. Predictions are made for the entire non-relativistic
energy domain. The phenomenon of dynamical suppression of
correlation effects in the ionization cross section is discussed.
\end{abstract}
\pacs{32.80.Fb, 32.80.-t, 31.25.Eb}
\maketitle

{\bf 1.} The single ionization of light atomic systems by photon
impact is one of the fundamental processes, which is being
persistently investigated during last decades
\cite{1,2,3,4,5,6,7,8,9,10,11,12}. Although the non-relativistic
problem for single photoeffect on H-like atom was solved
analytically by M. Stobbe already in 1930 \cite{13}, the further
generalization on the case of two-electron targets, such as, for
example, neutral He atom, is a non-trivial task. Since the strength
of the electron-electron interaction is comparable to that of the
electron-nucleus interaction, the single-particle treatment cannot
be appropriate. For theoretical calculations of ionization cross
sections, it is usual to employ sophisticated numerical methods
dealing with highly correlated wave functions. This allows one to
take into account the electron correlations beyond the
independent-particle approximation.

In this Letter, we evaluate the dominant contribution of correlation
effects to the cross section for single K-shell ionization of
He-like ions using the non-relativistic perturbation theory. To
zeroth-order approximation, it is assumed that the atomic nucleus is
the external source of the Coulomb field, while the interaction of
electrons with each other is neglected. The non-interacting
electrons are described by the Coulomb wave functions for the
discrete and continuous spectra (Furry picture). The
electron-electron interaction is treated within the framework of
perturbation theory. The latter exhibits fast convergence even for
small values of nuclear charge $Z$ \cite{14}. To first order of
perturbation theory, the problem of single ionization of He-like
ions is reduced to the evaluation of one-photon exchange diagrams.
The correlation corrections to the cross section arise due to
modification of the binding energy and wave functions of the initial
and final states. Using this approach, we have already deduced the
universal scaling behavior for the ionization cross section at high
photon energies \cite{15}. The formula obtained previously has a
simple analytical form, but it is valid only in the asymptotic
non-relativistic limit. This Letter reports on the extension of
results of our work \cite{15} on the entire non-relativistic energy
domain.

{\bf 2.} The non-relativistic problem of single ionization of a
K-shell bound electron by photon impact involves the following
quantities: the momentum $\bm{k}$, the energy $\omega=|\bm{k}|=k$,
and the polarization vector $\bm{\mathrm{e}}$ of an incident photon,
the binding energy $I=\eta^2/(2m)$ and the average momentum
$\eta=m\alpha Z$ of a K-shell electron, where $m$ is the electron
mass and $\alpha$ is the fine-structure constant ($\hbar=1$, $c=1$).
The outgoing electron is characterized by the energy $E_p
=\bm{p}^2/(2 m)$ and the momentum $\bm{p}$ at infinity. We shall
employ the Coulomb gauge, in which $(\bm{\mathrm e}\cdot \bm{k}) =
0$ and $(\bm{\mathrm e}^*\cdot \bm{\mathrm e}) = 1$. The parameter
$\alpha Z$ is supposed to be sufficiently small, that is, $\alpha Z
\ll 1$.

\begin{figure}[h]
\centerline{\includegraphics[scale=0.6]{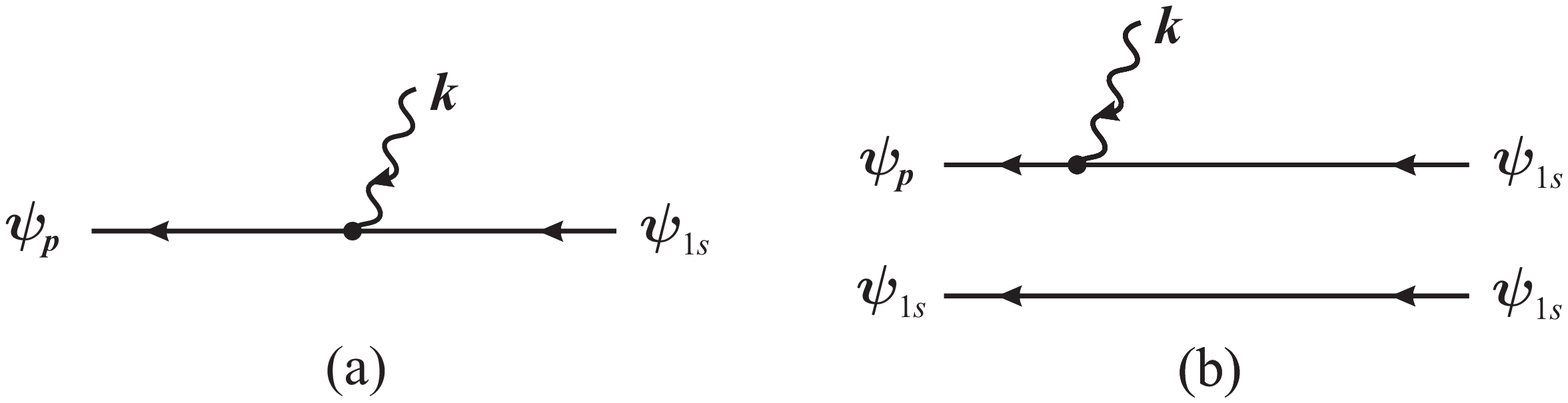}}
\caption{\label{fig1} Feynman diagrams for the single ionization of
a K-shell electron by a single photon. Diagram (a), hydrogen-like
ion; diagram (b), helium-like ion without taking into account the
electron-electron interaction.}
\end{figure}

The amplitude for single photoeffect on a bound K-shell electron is
described by the Feynman diagram depicted in Fig.~\ref{fig1}(a). The
energy-conservation law reads $E_p = \omega - I$. Accordingly, the
non-relativistic photoionization can proceed at the photon energies
$I \leqslant \omega \ll m$. Using the one-electron Coulomb wave
functions for the discrete and continuous spectra, Stobbe has
obtained an analytical expression for the cross section of single
K-shell photoeffect, which can be cast into the following form
\cite{13,16}
\begin{equation}
\sigma^+_{\mathrm{K}} = \alpha a_0^2
\frac{2^{9}\pi^2}{3Z^{ 2}}
\frac{\exp(-4\xi \cot^{-1}\xi )}{ (1+\xi^{-2})^4
[1 - \exp (-2\pi \xi)]} \,   ,
\label{eq1}                                           
\end{equation}
where $a_0 = 1/(m \alpha)$ is the Bohr radius. In Eq.~\eqref{eq1},
we have introduced the dimensionless parameter $\xi=\eta/p$. The
quantity $\xi^{-1}$ has the meaning of the momentum $p$ of the
outgoing electron, which is calibrated in units of the
characteristic momentum $\eta$. Due to the energy-conservation law,
$\xi$ is related with the dimensionless energy of the photon
$\varepsilon_{\gamma}= \omega/I$ according to $\xi =
1/\sqrt{\varepsilon_{\gamma} -1}$ or, equivalently, $\xi =
1/\sqrt{\varepsilon_{p}}$, where $\varepsilon_{p}=E_p/I$ is the
dimensionless energy of the outgoing electron. Equation \eqref{eq1}
is valid in the entire non-relativistic energy domain $1 \leqslant
\varepsilon_{\gamma} \ll 2(\alpha Z)^{-2}$. In the derivation, the
dipole approximation has been employed, which implies that the
photon momentum $\bm{k}$ is negligible compared with the electron
momentum $\bm{p}$.

In the non-relativistic photoeffect, one is often interested in the
asymptotic energy domain far beyond the threshold, which is
characterized by $1 \ll \varepsilon_{\gamma} \ll 2(\alpha Z)^{-2}$.
In this case, Eq.~\eqref{eq1} can be expanded with respect to the
small parameter $\xi \ll 1$. The leading term in the $\xi$
expansion is given by \cite{16}
\begin{equation}
\sigma^+_{\mathrm{K}} = \alpha a_0^2 \frac{2^{8}\pi}{3Z^{ 2}}\,
\varepsilon_{\gamma}^{-7/2}  \,  , \quad (\xi \ll 1) \,  .
\label{eq2}                                           
\end{equation}
In fact, Eq.~\eqref{eq2} is obtained using the Born approximation,
which describes the wave function of the outgoing electron in terms
of a plane wave. Since Eq.~\eqref{eq1} involves also the parameter
$\pi\xi$, which originates from the normalization factor of the
Coulomb wave function of the continuous spectrum, the convergence of
the $\xi$ expansion is sufficiently slow.

Another limit of expression \eqref{eq1} corresponds to the
ionization threshold ($\varepsilon_{\gamma} \to 1$). In this case,
the cross section reaches the constant value \cite{16}
\begin{equation}
\sigma^+_{\mathrm{K}} = \alpha a_0^2 \frac{2^{9}\pi^2}{3 e^4 Z^{2} }
\,  ,  \quad (\xi \gg 1) \,  ,
\label{eq3}                                           
\end{equation}
where $e\simeq 2.71828$ is the Napier-Euler number.

{\bf 3.}  To zeroth-order approximation, the amplitude
$\mathcal{A}^{(0)}$ for the single photoeffect on two-electron
atomic system in the ground state is described by the Feynman graph
depicted in Fig.~\ref{fig1}(b). In the momentum representation, it
can be written as follows
\begin{eqnarray}
\mathcal{A}^{(0)} &=& \sqrt{2} C  \frac{\exp(-2\xi \cot^{-1}\xi
)}{(1+\xi^{2})^2}  \,   , \label{eq4}\\
C &=& 8 \pi \eta N_{\gamma} N_{1s} N_{p}  (1 -
i\xi)\frac{(\bm{\mathrm{e}}\cdot \bm{p} )}{p^4}  \,   ,  \label{eq5}\\
N^2_{\gamma} &=& \frac{4\pi\alpha}{\eta^2 \varepsilon_{\gamma}m} \,
, \quad  N^2_{1s} = \frac{\eta^3}{\pi} \, , \nonumber\\
\quad N^2_{p} &=& \frac{2\pi\xi} {1 - \exp (-2\pi \xi)} \,  .
\nonumber
\end{eqnarray}
Here $\xi=\eta/p$ and the dipole approximation $(k=0)$ has been
again used in the derivation. The corresponding expression for the
total cross section $\sigma^+_{0}$ of the process reads
\begin{equation}
\sigma^+_{0} = 2\sigma_{\mathrm{K}}^+ \,  , \label{eq6}
\end{equation}
where $\sigma_{\mathrm{K}}^+$ is given by Eq.~\eqref{eq1}. The
factor 2 accounts for the number of electrons in the target. The
energy-conservation law keeps the same form as for the H-like ion,
namely, $\varepsilon_{p}= \varepsilon_{\gamma} - 1$. Equation
\eqref{eq6} is valid in the entire non-relativistic domain $1
\leqslant \varepsilon_{\gamma} \ll 2(\alpha Z)^{-2}$. However, in
the case of light two-electron atoms, the expression \eqref{eq6} is
a quite rough approximation, because the electron-electron
interaction is completely neglected. For example, for the neutral He
atom, Eq.~\eqref{eq6} predicts  $\sigma^+_{0} = 3.15$ Mb at the
ionization threshold ($\xi \to \infty$). The corresponding
experimental value turns out to be $\sigma^+_{\mathrm{exp}} = 7.40$
Mb at the photon energy $\omega = 24.59$ eV \cite{7}. The
significant discrepancy between these results is due to correlation
effects. Near the threshold, the approximation of non-interacting
electrons strongly underestimates the experimental cross section. On
the contrary, in the asymptotic high-energy limit, Eq.~\eqref{eq6}
yields an upper bound for the photoionization cross section
\cite{15}.

\begin{figure}[h]
\centerline{\includegraphics[scale=0.6]{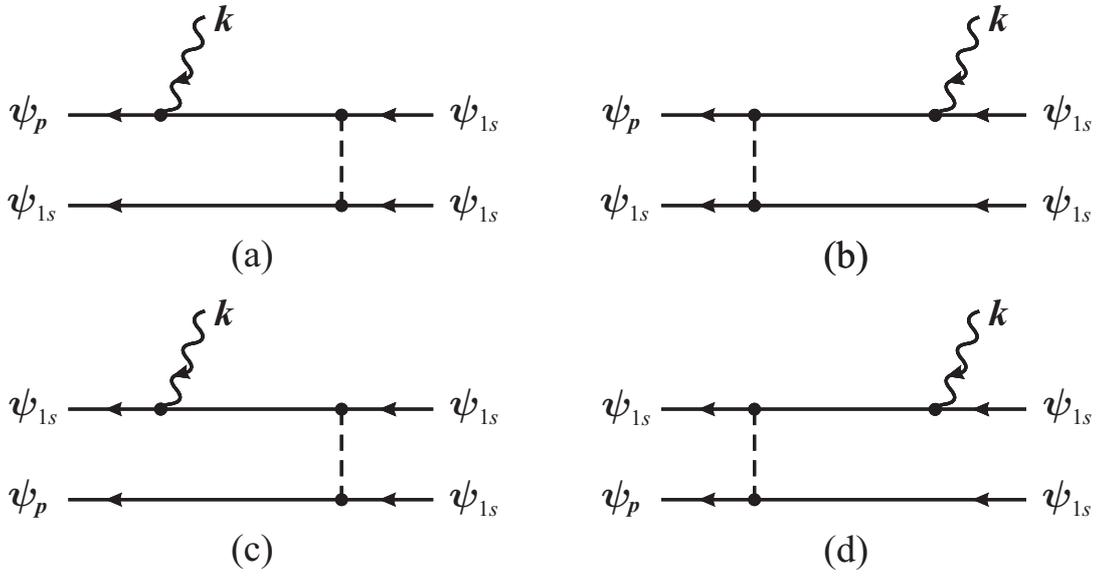}}
\caption{\label{fig2} Feynman diagrams for the single K-shell
photoionization of He-like ion. Diagrams (a) and (c) take into
account the electron-electron interaction in the initial state,
while diagrams (b) and (d) account for it in the final state.
The individual contributions of each diagram are gauge
dependent, while total contribution of all diagrams is gauge
independent.}
\end{figure}

Now we shall take into account the electron-electron interaction
using the perturbation theory. To first order, the amplitude
$\mathcal{A}^{(1)}$ for the single K-shell photoeffect on He-like
ion is described by the gauge invariant set of four Feynman graphs
depicted in Fig.~\ref{fig2}. The total amplitude for the process is
given by $\mathcal{A} \simeq \mathcal{A}^{(0)} + \mathcal{A}^{(1)}$,
where the amplitude \eqref{eq4} should be calculated with taking
into account the correlation correction to the binding energy. More
specifically, the energy-conservation law now implies
$\varepsilon_{p}= \varepsilon_{\gamma} - 1 + \delta_1$, where
$\delta_1=\Delta_1/Z$ is the correction to the binding energy of the
ground state calculated to first order of the perturbation theory
and $\Delta_1=5/4$ \cite{17}. In calculations of the amplitude
$\mathcal{A}^{(1)}$, the correlation correction $\delta_1$ to the
binding energy can be omitted, because it exceeds the desired level
of accuracy.

Note, that within the asymptotic high-energy domain, the binding
energies are assumed to be negligibly small with respect to the
photon energies. In addition, the wave function of the outgoing
electron can be approximated by the plane wave, so that the Feynman
graph in Fig.~\ref{fig2}(a) yields the dominant contribution to the
total cross section for single photoionization, provided the Coulomb
gauge is employed \cite{15}. However, for low energies, all Feynman
diagrams depicted in Fig.~\ref{fig2} are expected to contribute with
the same order of magnitude, while the plane-wave approximation
becomes inadequate.

Since the general method for evaluation of the Coulomb matrix
elements has been already described in more details \cite{18}, we
focus here on presentation of the explicit expression for the
amplitude $\mathcal{A}^{(1)} =\sum\limits_{\beta}
\mathcal{A}^{(1)}_{\beta}$ only. The individual contributions of
each diagram in Figs.~\ref{fig2}(a)--(d) read
\begin{eqnarray}
\mathcal{A}^{(1)}_{\mathrm{a}} &=& -\frac{\sqrt{2}}{8}  \frac{C}{Z}
\biggl\{ \left( \frac{10}{1+\xi^2} + \frac{3}{4} - 3 \ln 2 \right)
f(0) +  \nonumber\\
&&+ \left( \frac{1}{2} + 3 \ln 2 \right)f(1/2) + 3
\int_{0}^{1/2} \!\! dx f^{\prime}(x) \ln x  \biggr\} \,  , \label{eq7}\\
\mathcal{A}^{(1)}_{\mathrm{b}} &=& - i \sqrt{2} \, \xi  \frac{C}{Z}
\int\limits_{0}^{1} \frac{dx}{\Lambda} x^{1-i\xi} \left( \frac{1
+\Lambda}{1+ i\xi} \right)^{2i\xi}\left\{ \Phi_{\lambda} -
\Phi_{\nu} + \xi \frac{\partial \Phi_{\nu} }{\partial
\nu}\right\}{\!}_{\left|{} \atop {{\nu=2\xi}  \atop {\lambda \to 0}}
\right.}   \,  , \label{eq8}\\
\mathcal{A}^{(1)}_{\mathrm{c}} &=& 8\sqrt{2}\, \xi^4 \frac{C}{Z}
\int\limits_{0}^{1} \frac{dx}{u} \frac{x^{1-\xi/\mu}}{(u + \xi)^2}
\left(\frac{u + \mu }{\xi + \mu}\right)^{2\xi/\mu} \times\nonumber\\
&& \times \int\limits_{0}^{1} \frac{dy}{\chi^4} \sqrt{y} \, \left\{
\frac{\xi}{2} \chi + v(\xi + 2v)(2\xi + 3v)\right\} e^{-2\xi
\cot^{-1}(\xi + v )} \,  ,  \label{eq9}\\
\mathcal{A}^{(1)}_{\mathrm{d}} &=& 2\sqrt{2}\, \xi^4 \frac{C}{Z}
\frac{\partial^2}{\partial\zeta \partial\rho}\int\limits_{0}^{1}
\frac{dx}{\Lambda} x^{1-i\xi} \left( \frac{1 +\Lambda}{1+ i\xi}
\right)^{2i\xi} \times\nonumber\\
&&\times \int\limits_{0}^{1} \frac{dy \, y }{Q}  \frac{\left(1 - (Q
+i\rho)^2\right)^{i\xi-2}}{\left( -(1 +
Q + i\rho)^2 \right)^{i\xi}}{}_{\left|{} \atop { {\zeta=\xi}
\atop {\rho = \xi}} \right.} \,  ,  \label{eq10}\\
f(x) &=& \frac{(1-x^2)\exp[-2\xi\cot^{-1}(\tau \xi) ] } {[(1-x)^2
+\xi^2(1+x)^2]^2} \, , \quad  \tau= \frac{1+x}{1-x} \,  ,  \nonumber\\
\Phi_{\lambda} &=& \frac{\left(1 - (\Lambda +
i\lambda)^2\right)^{i\xi-2}}{ \left(- (1+ \Lambda +
i\lambda)^2\right)^{i\xi}} \,  ,  \quad
\Lambda = \sqrt{1 - x(1+ \xi^2)} \,  ,  \nonumber\\
u &=& \sqrt{1 -x + \xi^2 (2-x)} \,  ,  \quad  v= \sqrt{y}\, (u +
\xi) \,  , \quad  \mu = \sqrt{1 + 2\xi^2} \,   ,  \nonumber\\
\chi &=& 1 +(\xi + v)^2 \,  , \quad Q=\sqrt{y}\, (\Lambda + i\zeta) \,
. \nonumber
\end{eqnarray}
Here the function $C$ is defined by Eq.~\eqref{eq5}. In the integral
of Eq.~\eqref{eq7}, the prime at the function $f(x)$ denotes the
derivative with respect to the variable $x$. We note, that the
matrix element corresponding to the graph in Fig.~\ref{fig2}(a)
involves the reduced Green's function. This requires isolation and
cancelation of the pole term, which has been done analytically. In
Eq.~\eqref{eq8}, after taking the derivative with respect to $\nu$,
one should set $\nu=2 \xi$, where $\xi=\eta/p$. In addition, the
parameter $\lambda$ of a Yukawa-type screened Coulomb interaction
introduced for regularization of the integral should tend to zero.
In Eq.~\eqref{eq10}, after taking the derivatives with respect to
$\zeta$ and $\rho$, one should set $\zeta= \rho= \xi$. Note also
that the contribution of the graph in Fig.~\ref{fig2}(b) describing
the final-state interaction, contains a logarithmically divergent
term $\ln (\lambda/2p)$ for $\lambda \to 0$. However, this term
drops out in the expression for cross section of the process. This
fact reflects a general property of the wave function for continuous
spectrum, when it is constructed using perturbation theory with
respect to the Coulomb interaction. More precisely, the
logarithmically divergent terms of the wave function can be summed
up to an overall phase factor $\exp[i\xi\ln (\lambda/2p)]$, which
does not contribute to the cross section \cite{19,20,21,22,23}.

\begin{figure}[h]
\centerline{\includegraphics[scale=1.3]{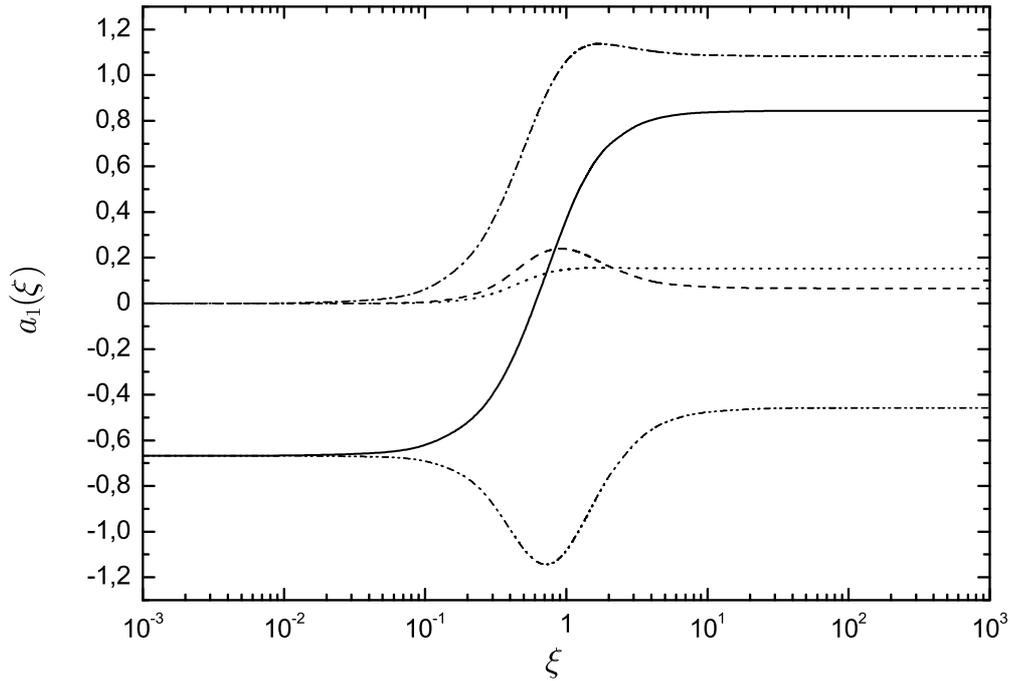}}
\caption{\label{fig3} Different contributions to the universal
function $a_1(\xi)$ calculated in Coulomb gauge. Dash-double dotted
line, contribution due to the diagram in Fig.~\ref{fig2}(a);
dash-dotted line, contribution due to the diagram in
Fig.~\ref{fig2}(b); dotted line, contribution due to the diagram in
Fig.~\ref{fig2}(c); dashed line, contribution due to the diagram in
Fig.~\ref{fig2}(d); solid line, total contribution of all diagrams.}
\end{figure}

Separating explicitly the contributions due correlation corrections
to the binding energy and wave functions, the cross section for
single K-shell photoionization of He-like ions can be cast into the
following form
\begin{equation}
\sigma^+ = \sigma_{0}^+ \left\{ 1 +
\left[\frac{\Delta_1}{\varepsilon_{\gamma}} + a_1(\xi)
 \right] \frac{1}{Z} \right\} \,  , \label{eq11}
\end{equation}
where $\sigma_{0}^+$ is given by Eq.~\eqref{eq6}, $\xi
=1/\sqrt{\varepsilon_{p}}$, $\varepsilon_{p}= \varepsilon_{\gamma} -
1 + \delta_1$, $\delta_1=\Delta_1/Z$, and $\Delta_1=5/4$. The
quantity $a_1(\xi)= 2Z \sum\limits_{\beta}\mathrm{Re} \left(
\mathcal{A}^{(1)}_{\beta}/ \mathcal{A}^{(0)}\right)$ is a universal
function of the parameter $\xi$ (see Fig.~\ref{fig3} and
Table~\ref{table1}). Equation \eqref{eq11} is valid in entire
non-relativistic domain $1 - \delta_1 \leqslant \varepsilon_{\gamma}
\ll 2(\alpha Z)^{-2}$ and represents the main result of this Letter.

The function $a_1(\xi)$ describes the ``dynamical correlation",
originating from the correlated two-electron wave functions both in
the initial and final states. Three characteristic domains can be
distinguished here, namely, the threshold domain $(\xi \gg 1)$, the
transition domain $(\xi \simeq 1)$, and the high-energy domain $(\xi
\ll 1)$. As can be seen from Fig.~\ref{fig3}, for values $\xi
\lesssim 10^{-1}$ and $\xi \gtrsim 10$, the function $a_1(\xi)$
saturates rapidly approaching the constant limits, while it
undergoes significant changes within the range $10^{-1}  \lesssim
\xi \lesssim 10$. Within the threshold domain, the consistent
account of the correlation interaction both in the initial and final
states becomes especially crucial. In the asymptotic high-energy
limit, $a_1(\xi)$ tends to the analytical value \cite{15}
\begin{equation}
a_1= -\frac{19}{16} + \frac{3}{4}\ln 2 \simeq -0.6676 \,  ,
\label{eq12}
\end{equation}
while the correlation correction in the binding energy is negligible
for the cross section. In this case, the correlation effect results
predominantly from the wave functions. In particular, within the
Coulomb gauge, this is just the contribution from the wave function
of the initial state. At about $\xi \simeq 0.6036$, the function
$a_1(\xi)$ tends to zero. Accordingly, the correlation effect in the
cross section arises due to the binding energy.

For some values of the parameter $\xi$, the correlation term in the
square brackets in Eq.~\eqref{eq11} vanishes. In this case, the
correlation effect due to the binding energy cancels with that due
to the wave functions. As a result, the cross section \eqref{eq11}
coincides with the Coulomb prediction \eqref{eq6}. In
Table~\ref{table2}, we consider the particular examples, where the
dynamical suppression of correlation effects take place. Note, that
for neutral helium atom the interpolation of experimental cross
sections predicted in Ref.~\cite{7}  yields $\sigma^+_{\mathrm{exp}}
=14.2$ kb at the photon energy $\omega=308.2$ eV. The deviation from
the corresponding theoretical value  $\sigma_0^+=15.6$ kb is
exclusively due to higher-order correlation corrections neglected in
the present consideration.

In Table~\ref{table3}, we present a comparison between our
predictions according to Eq.~\eqref{eq11} for several two-electron
targets at different photon energies with the numerical calculations
of the work \cite{2}. Bell and Kingston used the Hartree-Fock wave
functions for the continuum state and many-parametrical variational
wave functions for the ground state. The results of Ref.~\cite{2}
are slightly gauge dependent. For very light targets, such as
neutral He atom, the account for the correlation correction
$\delta_1$ to the binding energy is still not sufficient to
reproduce the experimental thresholds. Certainly, one needs to go
beyond the first-order approximation. However, although consistent
calculations of the binding energies have been performed with taking
into account of two- and three-photon exchange diagrams
\cite{24,25}, the evaluation of ionization cross sections at the
same level of accuracy seems to be not feasible in the near future.
Accordingly, in Table \ref{table3}, we merely employ the
experimental threshold energies $\omega$. As can be seen, the
formula \eqref{eq11} turns out to be in good agreement with the
numerical results by Bell and Kingston \cite{2} both at the
ionization threshold and beyond.

\acknowledgments

A.M. is grateful to the Dresden University of Technology for the
hospitality and for financial support from Max Planck Institute for
the Physics of Complex Systems. A.N. and G.P. acknowledge financial
support from DFG, BMBF, and GSI. This research has been also
supported by RFBR (Grant no. 05-02-16914) and INTAS (Grant no.
06-1000012-8881).

\newpage
\begin{table}
\caption{\label{table1} For various values of the dimensionless
parameter $\xi$, the universal quantities $a_1(\xi)$ are tabulated.
Numbers in parentheses indicate powers of 10.}

\begin{center}
\begin{tabular}{l l  l l  l l  l l  l l  l l  l l} \hline
\multicolumn{1}{c}{$\xi$} & \multicolumn{1}{c}{$a_1(\xi)$} &
\multicolumn{1}{c}{$\xi$} & \multicolumn{1}{c}{$a_1(\xi)$} &
\multicolumn{1}{c}{$\xi$} & \multicolumn{1}{c}{$a_1(\xi)$} &
\multicolumn{1}{c}{$\xi$} & \multicolumn{1}{c}{$a_1(\xi)$} &
\multicolumn{1}{c}{$\xi$} & \multicolumn{1}{c}{$a_1(\xi)$} &
\multicolumn{1}{c}{$\xi$} & \multicolumn{1}{c}{$a_1(\xi)$} &
\multicolumn{1}{c}{$\xi$} & \multicolumn{1}{c}{$a_1(\xi)$}
\\  \hline
$1.0(-3)$  & $-0.6676$ & $0.2$ &  $-0.5362$ &  $0.7$  &  $0.1099$ &
$1.2$ & $0.4822$ &  $1.7$ & $0.6448$  &  $4.0$  &  $0.8048$  &
$30$ & $0.8433$  \\
$0.5(-2)$&  $-0.6675$ & $0.3$ & $-0.4087$ & $0.8$ &  $0.2097$ &
$1.3$ & $0.5262$  & $1.8$   & $0.6645$  &  $5.0$  &  $0.8190$  &
$50$  & $0.8438$    \\
$1.0(-2)$ & $-0.6672$  & $0.4$ &  $-0.2693$ & $0.9$  & $0.2955$ &
$1.4$  & $0.5634$ & $1.9$  & $0.6816$   &  $7.0$  &  $0.8313$  & $80$ &
$0.8439$  \\
$0.5(-1)$ & $-0.6580$ & $0.5$ &  $-0.1319$  & $1.0$ & $0.3684$ &
$1.5$ & $0.5949$ & $2.0$ & $0.6965$ & $10$  &  $0.8378$  &  $1.0(2)$ &
$0.8439$   \\
$0.1$  & $-0.6309$ & $0.6$  &  $-0.0043$  &  $1.1$  & $0.4302$ &
$1.6$ & $0.6218$  & $3.0$  &$0.7763$ &  $20$  &  $0.8424$  & $1.0(3)$ &
$0.8440$   \\
\hline
\end{tabular}
\end{center}
\end{table}

\begin{table}
\caption{\label{table2} For different values of the nuclear charge
$Z$, the parameters $\xi$, the dimensionless energies
$\varepsilon_{\gamma}$, the Coulomb ionization potentials $I$, the
photon energies $\omega$, and the corresponding ionization cross
sections $\sigma^+$ are tabulated. For the considered values of $\xi$,
the predictions according to Eq.~\eqref{eq11} coincide with those
according to Eq.~\eqref{eq6}.}

\begin{center}
\begin{tabular}{l l  l ll  l  l  ll l  l l } \hline
\multicolumn{1}{c}{$Z$} & \multicolumn{1}{c}{$\xi$}
&\multicolumn{1}{c}{$\varepsilon_{\gamma}$} & \multicolumn{1}{c}{$I$
(eV)} & \multicolumn{1}{c}{$\omega$ (eV)} &
\multicolumn{1}{c}{$\sigma^+$ (kb)} & \multicolumn{1}{c}{$Z$} &
\multicolumn{1}{c}{$\xi$}
&\multicolumn{1}{c}{$\varepsilon_{\gamma}$} & \multicolumn{1}{c}{$I$
(eV)} & \multicolumn{1}{c}{$\omega$ (keV)} &
\multicolumn{1}{c}{$\sigma^+$ (kb)}   \\
\hline
 2   &  0.4348  &  5.663 & 54.42 & 308.2  & 15.6 & 7  & 0.4422  &  5.936  & 666.7 & 3.957 &  1.39  \\
 3   &  0.4383  &  5.789 & 122.4 & 708.8  & 7.24 & 8  & 0.4425  &  5.950  & 870.8 & 5.181 &  1.07  \\
 4   &  0.4400  &  5.853 & 217.7 & 1274   & 4.15 & 9  & 0.4428  &  5.961  & 1102  & 6.569 &  0.847  \\
 5   &  0.4410  &  5.891 & 340.1 & 2004   & 2.69 & 10 & 0.4430  &  5.970  & 1361  & 8.122 &  0.688  \\
 6   &  0.4417  &  5.917 & 489.8 & 2898   & 1.88 & 11 & 0.4432  &  5.977  & 1646  & 9.840 &  0.570  \\
\hline
\end{tabular}
\end{center}
\end{table}

\clearpage

\newpage
\begin{table}
\caption{\label{table3} For various two-electron targets, the
nuclear charges $Z$, the photon energies $\omega$, the corresponding
dimensionless energies $\varepsilon_{\gamma}$, and the
photoionization cross sections $\sigma^+$  are tabulated. Our
predictions are made according to Eq.~\eqref{eq11}. The calculations
by Bell and Kingston \cite{2} have been performed using different
gauges: velocity (a) and length (b). Numbers in parentheses indicate
powers of 10.}

\begin{center}
\begin{tabular}{l l  l ll  l l  ll l  l l } \hline
\multicolumn{1}{c}{Target} & \multicolumn{1}{c}{$Z$}
&\multicolumn{1}{c}{$\omega$ (eV)} & \multicolumn{1}{c}{$\varepsilon_{\gamma}$}
& \multicolumn{3}{c}{$\sigma^+$
(Mb)} & \multicolumn{1}{c}{$\omega$ (keV)} &
\multicolumn{1}{c}{$\varepsilon_{\gamma}$} &
\multicolumn{3}{c}{$\sigma^+$ (b)} \\
\cline{5-7} \cline{10-12} &  & & &
\multicolumn{1}{c}{Eq.~\eqref{eq11}} &
\multicolumn{1}{c}{Ref.~\cite{2}$^{\rm a}$}  &
\multicolumn{1}{c}{Ref.~\cite{2}$^{\rm b}$} &
&&\multicolumn{1}{c}{Eq.~\eqref{eq11}}
&\multicolumn{1}{c}{Ref.~\cite{2}$^{\rm a}$}
& \multicolumn{1}{c}{Ref.~\cite{2}$^{\rm b}$}   \\
\hline
He   & 2  & 24.59  & 0.452 & 7.19  & 7.15  & 7.32  & 0.456  & 8.375  & 4.68(3) & 4.40(3) & 4.80(3)   \\
     &    & 61.23  & 1.125 & 1.21  & 1.07  & 1.14  & 0.891  & 16.37 & 553 & 527 & 558  \\
     &    & 183.7  & 3.375 & $7.19(-2)$ & $6.64(-2)$ & $7.34(-2)$ & 1.327 & 24.37  &  151 & 146 & 151  \\
     &    & 238.1  & 4.375 & $3.39(-2)$ & $3.15(-2)$ & $3.47(-2)$ & 1.762  & 32.37  & 59.5  & 57.4  &  59.0  \\
     &    & 346.9  & 6.375 & $1.09(-2)$ & $1.02(-2)$ & $1.12(-2)$ & 3.503  & 64.37  & 6.05  & 5.74  & 6.12  \\
 \hline
Li$^{+}$  & 3  & 75.64  & 0.618 & 2.49       & 2.57       &  2.61      & 0.289  & 2.361 & 9.88(4) & 9.56(4) &  9.96(4)   \\
          &    & 112.2  & 0.917 & 1.07       & 1.02       &  1.05      & 0.398  & 3.250 & 4.01(4) & 3.88(4) &  4.06(4)  \\
          &    & 153.1  & 1.250 & $5.19(-1)$ & $4.98(-1)$ & $5.11(-1)$ & 1.378  & 11.25 & 918     & 897     &  928  \\
          &    & 180.3  & 1.472 & $3.45(-1)$ & $3.33(-1)$ & $3.43(-1)$ & 1.813  & 14.81 & 383     & 375     &  386  \\
          &    & 234.7  & 1.917 & $1.74(-1)$ & $1.68(-1)$ & $1.74(-1)$ & 3.554  & 29.03 & 43.2    & 41.7    &  44.0  \\
 \hline
B$^{3+}$  & 5  & 259.3  & 0.762  & 0.730  & 0.743 & 0.748 & 0.473  & 1.390  & 1.67(5) & 1.65(5) &  1.67(5)  \\
          &    & 295.9  & 0.870  & 0.537  & 0.528 & 0.533 & 0.582  & 1.710  & 9.65(4) & 9.53(4) &  9.64(4)  \\
          &    & 336.7  & 0.990  & 0.393  & 0.388 & 0.391 & 1.126  & 3.310  & 1.50(4) & 1.48(4) &  1.51(4)  \\
          &    & 363.9  & 1.070  & 0.325  & 0.321 & 0.323 & 1.561  & 4.590  & 5.72(3) & 5.64(3) &  5.74(3)   \\
          &    & 418.4  & 1.230  & 0.229  & 0.226 & 0.228 & 3.738  & 10.99  & 388  & 384   &  399   \\
 \hline
O$^{6+}$  & 8 & 739.0 & 0.849 & 0.251      & 0.253      & 0.254      & 3.782  & 4.344 & 2.77(3) & 2.76(3) & 2.78(3)  \\
          &   & 1497  & 1.719 & $3.99(-2)$ & $3.98(-2)$ & $4.00(-2)$ & 5.306  & 6.094 & 994     & 990  &  996 \\
          &   & 1878  & 2.156 & $2.13(-2)$ & $2.12(-2)$ & $2.13(-2)$ & 6.830  & 7.844 & 457  & 455  & 458  \\
          &   & 2259  & 2.594 & $1.26(-2)$ & $1.25(-2)$ & $1.26(-2)$ & 9.878  & 11.34 & 144  & 144  & 144  \\
          &   & 3020  & 3.469 & $5.40(-3)$ & $5.38(-3)$ & $5.41(-3)$ & 12.92  & 14.84 & 61.4 & 61.1 & 61.4   \\
 \hline
Ne$^{8+}$ & 10 & 1195  & 0.878 & 0.153   & 0.154 & 0.155 & 7.068  & 5.195 & 1.05(3) & 1.05(3) & 1.05(3)  \\
          &    & 1313  & 0.965 & 0.121   & 0.121 & 0.121 & 9.027  & 6.635 & 498 &  497  &  498  \\
          &    & 2660  & 1.955 & $1.83(-2)$ & $1.83(-2)$ &  $1.83(-2)$  & 12.95  & 9.515 & 163 & 162   &  163 \\
          &    & 4129  & 3.035 & $5.20(-3)$ & $5.19(-3)$ &  $5.21(-3)$  & 16.86  & 12.39 & 70.9 & 70.7 &  70.9 \\
          &    & 5109  & 3.755 & $2.78(-3)$ & $2.77(-3)$ &  $2.79(-3)$  & 24.46  & 17.97 & 21.7 & 21.4 &  21.5 \\
\hline
\end{tabular}
\end{center}
\end{table}

\clearpage

\newpage

\end{document}